\documentclass[final,5p,times,twocolumn,authoryear]{elsarticle}

\usepackage{amssymb}
\usepackage{siunitx}
\usepackage{enumitem}
\usepackage{lipsum}

\journal{International Journal of Human-Computer Studies}

\begin{document}

\begin{frontmatter}



\title{From Driver to Supervisor: Comparing Cognitive Load and EEG-based Attentional Resource Allocation across Automation Levels}


\author[inst1]{Nikol Figalová}
\author[inst2]{Hans-Joachim Bieg}
\author[inst3]{Julian Elias Reiser}
\author[inst4]{Yuan-Cheng Liu}
\author[inst5]{Martin Baumann}
\author[inst6]{Lewis Chuang}
\author[inst1]{Olga Pollatos}

\affiliation[inst1]{organization={Ulm University, Institute of Psychology and Education, Department of Clinical and Health Psychology},
            addressline={Albert-Einstein-Allee 45}, 
            city={Ulm},
            postcode={89081}, 
            country={Germany}}

\affiliation[inst2]{organization={Robert Bosch GmbH, Corporate Research},
            addressline={Robert-Bosch-Campus 1}, 
            city={Renningen},
            postcode={71272}, 
            country={Germany}}

\affiliation[inst3]{organization={Leibniz Research Centre for Working Environment and Human Factors, Ergonomics Department},
            addressline={Ardeystraße 67}, 
            city={Dortmund},
            postcode={44139}, 
            country={Germany}}
            
\affiliation[inst4]{organization={Chair of Ergonomics, Department of Mechanical Engineering, School of Engineering and Design},
            addressline={Boltzmannstraße 15}, 
            city={Garching bei München},
            postcode={85748}, 
            country={Germany}}    

\affiliation[inst5]{organization={Ulm University, Institute of Psychology and Education, Department of Human Factors},
            addressline={Albert-Einstein-Allee 45}, 
            city={Ulm},
            postcode={89081}, 
            country={Germany}}

\affiliation[inst6]{organization={Chemnitz University of Technology, Faculty of Humanities, Institute for Media Research, Department of Human and Technology},
            addressline={}, 
            city={Chemnitz},
            postcode={09107}, 
            country={Germany}}

\begin{abstract}
With increasing automation, drivers' roles transition from active operators to passive system supervisors, affecting their behaviour and cognitive processes. This study addresses the attentional resource allocation and subjective cognitive load during manual, SAE Level 2, and SAE Level 3 driving in a realistic environment. An experiment was conducted on a test track with 30 participants using a prototype automated vehicle. While driving, participants were subjected to a passive auditory oddball task and their electroencephalogram was recorded. The study analysed the amplitude of the P3a event-related potential component elicited by novel environmental stimuli, an objective measure of attentional resource allocation. The subjective cognitive load was assessed using the NASA Task Load Index. Results showed no significant difference in subjective cognitive load between manual and Level 2 driving, but a decrease in subjective cognitive load in Level 3 driving. The P3a amplitude was highest during manual driving, indicating increased attentional resource allocation to environmental sounds compared to Level 2 and Level 3 driving. This may suggest that during automated driving, drivers allocate fewer attentional resources to processing environmental information. It remains unclear whether the decreased processing of environmental stimuli in automated driving is due to top-down attention control (leading to attention withdrawal) or bottom-up competition for resources induced by cognitive load. This study provides novel empirical evidence on resource allocation and subjective cognitive load in automated driving. The findings highlight the importance of managing drivers' attention and cognitive load with implications for enhancing automation safety and the design of user interfaces.
\end{abstract}




\end{frontmatter}

\section{Introduction}
Until fully automated vehicles (AVs) are available, humans remain involved in operating partially (SAE Level 2; or L2) and conditionally (SAE Level 3; or L3) AVs \citep{sae2016j3016}. Nevertheless, with increasing automation, the drivers' roles change from primary operators to system supervisors \citep{bainbridge1983ironies}. The progressively more passive role could, however, lead to mental underload \citep{stapel2019automated, mcwilliams2021underload}, passive fatigue \citep{matthews2019dangerous, figalova2023fatigue}, and mind-wandering \citep{galera2012mind, baldwin2017detecting}. The disengagement with the driving task can result in drivers' inability to maintain alertness during automated driving \citep{vogelpohl2019asleep}. Nevertheless, drivers of L2 and L3 AVs remain responsible for the vehicle's safety \citep{sae2016j3016}. 

To optimise driver-vehicle interaction and ensure traffic safety, we must design user-centred AV systems that cooperatively interact with drivers and adapt to their current state and needs. To design such AV systems, we need to understand the cognitive processes underlying driver-vehicle interaction and explore the differences between different levels of automation. Although this topic had been previously addressed by the scientific community, most literature reports on data obtained in a driving simulator using subjective, self-report measures only. However, other researchers report discrepancies in a simulator and real-world findings when it comes to, for example, speed of driving \citep{zoller2019driver}, lateral position \citep{daurat2013lorazepam}, eye fixations \citep{fors2013simulator}, heart rate \citep{johnson2011physiological}, or mental workload \citep{mueller2015driving} (for a review, see \citet{wynne2019systematic}). Moreover, other researchers also report discrepancies between subjective and objective measures of cognitive states \citep{stapel2019automated, stapel2017driver, large2017exploring}. Additionally, we found no empirical studies which directly compare different levels of automation. 

To address these limitations and contribute to a more comprehensive understanding of the cognitive processes involved in driver-vehicle interaction during AV operation, this study aimed to compare L2 and L3 driving with manual driving in a realistic environment using both objective and subjective measures of cognitive load. Specifically, we investigated attentional resource allocation and self-reported cognitive load of drivers who operated SAE L2 and L3 vehicles and compared it to manual driving. We conducted an empirical test track experiment using an AV prototype and electroencephalography (EEG) to measure neural activity. The self-reported cognitive load was assessed using the NASA Task Load Index (NASA-TLX; \cite{hart2006nasa}), and attentional resource allocation was quantified using the amplitude of the P3a event-related potential (ERP) evoked by novel environmental sounds \citep{baldwin2017detecting}.

To our best knowledge, this is the first attempt to compare L2 and L3 driving with manual driving in a realistic environment using both objective and subjective measures of cognitive load. By investigating the neural mechanisms of drivers' behaviour during AV operation, our study provides insights into how the brain processes information during AV operation and how it differs from manual driving. Ultimately, our results may help design intelligent user-centred interfaces for future automated vehicles.

\section{Related Work}
\subsection{How Vehicle Automation Changes Drivers Roles}\label{bla}

Manual driving is a complex task that requires drivers to be aware of stimuli originating from different distances, directions, and sources, as a critical event can arise from any direction. Visual perception is crucial for processing important visual information from the environment and maintaining situational awareness. Auditory perception allows drivers to be aware of important auditory cues, such as horns or sirens, which can indicate potential hazards. Safe manual driving requires drivers to integrate information from multiple sensory modalities to form a unified perception of the world \citep{spence2020crossmodal}. It also requires drivers to update their perceptual-motor loop continuously and remain calibrated to the vehicle dynamics \citep{mole2019getting}.

Automated driving, however, places different demands on drivers. SAE L2 AVs can perform longitudinal and lateral control but require drivers to continuously supervise the AV system and immediately respond to automation failures. SAE L3 AVs allow drivers to engage in non-driving-related activities and only respond to requests to intervene once the AV reaches its operational boundary \citep{sae2016j3016}. As automation increases, the vehicle performs more active tasks, such as steering and navigating complex traffic situations, with drivers primarily serving a supervisory role. This shift in the role could disrupt the perceptual-motor loop \citep{mole2019getting} and decrease situational awareness \citep{merat2009bdrivers}. 

The role of drivers changes with increasing automation, which could lead to drivers being unprepared for the shift in task demands \citep{brandenburg2014switching, merat2009drivers}. This change in demands could jeopardize drivers' attention \citep{vogelpohl2019asleep, merat2014transition} and their ability to incorporate information from the environment \citep{galera2012mind, scheer2018auditory}. 

Many researchers addressed the changing role of the driver and its implications on how drivers interact with the automated system. \citet{jamson2013behavioural} studied the behavioural changes in drivers when experiencing highly automated driving. They found that drivers became more involved with in-vehicle entertainment tasks than in manual driving, impacting how much attention they pay to the environment. \citet{frison2019you} found that interacting with automated systems might be tedious and unstimulating, which leads to low psychological fulfilment and an unsatisfied user experience. \citet{biondi2019human} argue for collaborative human-vehicle interactions. 

To design AV systems that support drivers in their new role, we must understand their experiences when operating an AV and how those experiences differ at different levels of automation. This understanding will help us design user-centred interfaces for automated vehicles that better support drivers in their new roles.

\subsection{Attention, Resource Allocation, and Cognitive Load in Automated Driving}
Attention is a fundamental characteristic of all perceptual and cognitive processes \citep{chun2011taxonomy}. It can be classified as external (bottom-up) or internal (top-down). Bottom-up attention is reflexively triggered by environmental stimuli based on the saliency of a stimulus, whereas top-down attention involves directing attention based on cognitive goals. Driving scenarios present a dynamic interplay between bottom-up and top-down attention control \citep{spence2020crossmodal}. An example of bottom-up attention occurs when a red brake light of a car in front lights up, drawing visual attention and triggering the driver's bottom-up attention. This visual stimulus triggers an automatic, bottom-up attention response. 

Conversely, top-down attention control can be seen when the driver purposefully attends to traffic signs along the road. Their knowledge of the signs' importance, along with a commitment to driving safely and according to law, directs their attention towards the traffic signs amidst various distractions. This proactive, goal-driven behaviour demonstrates top-down attention control. 

The human brain has limited attentional capacity; therefore, information must compete for processing resources. Attention is then allocated to the most behaviorally relevant events \citep{chun2011taxonomy}. The saliency of stimuli is the primary mechanism for selection, but it can be modified via top-down control \citep{wang2020salience}. Several areas of the prefrontal and parietal cortex (for a review, see \citet{baluch2011mechanisms}) are involved in setting up these perceptual filters. The result is selection biased by cognitive goals, which means drivers can decide how much of their attention capacity is used for the driving task. 

Attention and the allocation of processing resources are closely related to cognitive load, which refers to the cognitive resources required to complete a task. When the cognitive load induced by a task is high, the processing resources are employed to complete the task. Conversely, during low-demanding tasks, there are enough free processing resources available. The effect of vehicle automation on cognitive load has been extensively studied in driving simulators (for an overview, see \cite{de2014effects}). However, the literature suggests that the simulated environment is limited in reproducing real traffic conditions \citep{groh2019simulation, hock2018design} and some authors report a discrepancy between a simulator and real-world results. For example, other studies report discrepancies in a simulator and real-world findings in regards to the speed of driving \citep{zoller2019driver}, lateral position \citep{daurat2013lorazepam}, eye fixations \citep{fors2013simulator}, heart rate \citep{johnson2011physiological}, or mental workload \citep{mueller2015driving} (for a review, see \citet{wynne2019systematic}). Therefore, driving simulator findings might have limited relevance in the real world.

Few real-world studies addressing cognitive load in automated driving have been conducted, and the authors report contradictory results. For example, \citet{biondi201880} reported a reduction in drivers' objective cognitive load (indexed by heart rate variability) in L2 driving compared to manual driving, while \citet{mcdonnell2021your} found no change in objective cognitive load (indexed by alpha and theta power) between manual and L2 driving. Contradicting results were reported by \citet{kim2023partially}, who found that L2 drivers experienced higher self-reported (measured by NASA-TLX) cognitive load in L2 compared to manual driving. Focusing on L3 driving, \citet{varhelyi2021driving} reported no differences in self-reported cognitive load (measured by a raw task-load index \citep{byers1989traditional}) between L3 and manual driving. In contrast, \citet{banks2016keep} concluded that L3 AV drivers perceive increased self-reported cognitive load (measured by NASA-TLX) compared to manual driving. Some authors also report discrepancies between self-report and objective measures of cognitive load \citep{stapel2019automated, stapel2017driver, large2017exploring}. These discrepancies, however, do not necessarily mean contradicting results. \citet{de1996measurement} suggest that different approaches to measuring cognitive load are sensitive to different components of cognitive load and recommends combining measures from different categories (self-report, performance-based, and physiological measures) to gain a comprehensive understanding of cognitive load. \citet{stapel2019automated} suggests that the discrepancy might occur because drivers tend to underestimate the actual cognitive load of passive supervision. 

Understanding drivers' attention, how they allocate processing resources, and the cognitive demands of automated driving is crucial for designing user-centred AV systems. However, many inconsistencies exist between self-report and objective measures of cognitive load (observed by, for example, \citet{stapel2019automated, stapel2017driver} or \citet{large2017exploring}) and between simulator and real-world findings (observed by, for example, \citet{fors2013simulator, johnson2011physiological, mueller2015driving}; for a review, see \citet{wynne2019systematic}). Accumulating more empirical evidence combining different approaches and methods is necessary for a comprehensive understanding of the problem and for developing neuroadaptive interfaces that allow the assessment of drivers' cognitive states in real time \citep{krol2016task}. Finally, utilising real-world studies over simulator experiments can help improve the factual relevance of the findings. 

\subsection{P3a as an Objective Measure of Resource Allocation}
Event-related potentials (ERPs) offer a noninvasive way to measure the brain's electrical activity, thereby providing valuable insights into cognitive and neural processes (for a comprehensive review, refer to \citet{ghani2020erp} or \citet{luck_2014}). The auditory P3a ERP component is of particular interest in neuroergonomics. The P3a component is typically associated with the brain's attentional response to novel or unexpected events or stimuli. It can be recorded over the frontocentral brain area and peaks approximately 300 ms after the onset of unexpected auditory stimuli. The P3a amplitude is related to the cognitive evaluation of the relevance of environmental stimuli and can be considered an index of involuntary attention switching \citep{escera2007role}. 

The involuntary attention switching is an information-processing cascade associated with auditory stimulus categorization and evaluation \citep{kok2001utility}. It occurs in three stages. In the perceptual stage, identified around 100-150 ms post-stimulus, the brain recognises the presence of the stimulus. This stage is characterised by the mismatch negativity (MMN) and the auditory N1 ERP components. Subsequently, in the categorisation stage, evident around 250-400 ms after introducing the stimulus, the brain cognitively evaluates the stimulus. This stage is reflected in the P3a ERP component. Once the distracting stimulus is categorised as irrelevant, an attention switch occurs. This switch is reflected by the reorienting negativity component (RON) with approximately 500-600 ms latency \citep{escera2007role}.

The P3a can be elicited by novel auditory stimuli within a passive oddball task paradigm. An oddball task is a common paradigm used in psychological research to study the perception, attention, and processing of stimuli \citep{polich2007updating}. 

In an active auditory oddball task, participants are presented with a sequence of auditory stimuli. The majority of these stimuli are the same, standard sounds. A small proportion are different, oddball stimuli. Traditionally, the participant's task is to detect the oddball stimuli and indicate their presence, e.g., by a button press \citep{parmentier2008cognitive}. In the passive oddball task, participants are presented with three types of stimuli: (1) frequent and (2) infrequent beep tones of distinct frequencies and (3) rare novel sounds. Contrary to the active oddball task, participants are instructed not to respond to the stimuli in the passive oddball task.

The passive oddball task can be presented alongside a primary task. While performing the primary task, participants are exposed to the sound presentation of the passive oddball task and their EEG is recorded. The amplitude of the P3a component elicited by the rare, novel sounds is influenced by the demands of the primary task \citep{escera1998neural}. A vital factor in this process is the competition for limited processing resources \citep{chun2011taxonomy}. In essence, as the primary task's demands increase, the pool of resources available for processing task-irrelevant auditory cues diminishes. This relationship presents an inverse correlation between the two \citep{harmony2000primary} - with fewer resources available to process the cues, the P3a amplitude is diminished. As a result, the P3a amplitude can be used to quantify the attentional resources allocated for stimulus processing; and therefore reflects the cognitive load of the primary task \citep{kramer1995assessment, sirevaag1993assessment, polich2007updating, figalova2023manipulating}.

However, treating the P3a amplitude as a direct measurement of the cognitive demands of the primary task might be misleading \citep{kok2001utility}. This direct relationship might be present in cognitively demanding tasks. In such situations, not enough resources are left for the processing of the task-irrelevant auditory cues. This situation leads to a competition for cognitive resources \citep{chun2011taxonomy}, subsequently resulting in a decreased P3a amplitude evoked by the task-irrelevant auditory cues \citep{polich2007updating}. However, in complex situations that require low to moderate cognitive load (such as driving), the saliency of the distractors might not be the only mechanism that determines the amount of processing resources used, and other factors might come into play. Top-down attention control appears to be one of these factors. 

The effect of top-down attention control on the P3a amplitude has been shown by \citet{cahn2009meditation}, who presented task-irrelevant auditory distractors to experienced meditators during meditation. They found decreased P3a amplitude evoked by the distracting stimuli and concluded that meditation reduced cognitive reactivity. This finding suggests the influence of top-down attention control on the processing of environmental auditory stimuli. This finding might have implications for neuroergonomic research because it suggests that the P3a amplitude is influenced not only by cognitive load. Even though meditation and driving are very different situations, the effect of top-down attention allocation might be present in both. 

The characteristics of the P3a ERP component have been previously applied in human-computer interaction research. The amplitude of the P3a has been used to assess the design and usability of auditory notifications in complex environments such as gaming \citep{lee2014eeg} or in-vehicle warning systems \citep{huang2019eeg}. \citet{cherng2018understanding} used the P3a ERP component to evaluate the impact of acoustic features of auditory notifications on awareness and attention shifting. Later, \citet{cherng2019measuring} also evaluated the influences of musical parameters on behavioural responses to audio notifications.

Moreover, several researchers studied the P3a amplitude in the context of cognitive load while driving. Specifically, 1) researchers compared the P3a amplitudes between stationary and driving conditions \citep{wester2008event}, 2) studied the effect of steering demands on P3a amplitude \citep{scheer2016steering}, 3) and compared the P3a amplitude in a stationary condition to manual and automated driving \citep{van2018susceptibility}. These studies found that the P3a amplitude was increased in the stationary condition compared to automated driving and that the P3a amplitude was reduced in manual driving compared to automated driving. Furthermore, \citet{van2021effect} investigated the effect of cognitive load on the processing of auditory cues during automated driving and found that the P3a amplitude was reduced when performing a task that induces cognitive load. These findings suggest that drivers allocate more attentional resources to process environmental stimuli when operating an automated vehicle than during manual driving. The results are then interpreted in light of limited attentional capacity and resource competition. However, none of the above-mentioned studies considers top-down attention control and its importance in tasks inducing lower cognitive load. In such situations, the competition for the limited pool of processing resources might not explain the observed changes in the P3a amplitude.

\subsection{Research Gaps}
To design AVs that are safe, efficient, and user-centred, it is crucial to get a comprehensive understanding of the drivers' experience when interacting with such vehicles. We must explore not only the perceived level (how drivers think they experience AVs) but also the implicit effects on their cognitive processes. Moreover, we must understand the differences between different levels of automation and their implications for AV design. Despite a substantial body of literature addressing some aspects of this issue, we identified several research gaps that must be addressed. 

Typically, such research is conducted in driving simulators, which affects the ecological validity of the findings due to the restricted real-world applicability. Furthermore, there is usually no comparison between different levels of automation, thus making it difficult to extrapolate differences in experiences across automation levels. Only a fraction of studies integrate both objective and subjective measurements of cognitive load. When these measures are combined, sometimes the results exhibit inconsistencies, indicating a need for further investigation to enhance our comprehension. Lastly, studies employing the passive oddball task generally interpret results from the standpoint of resource competition but tend to overlook the potential significance of top-down attentional control in situations necessitating low to medium cognitive load, such as (automated) driving.

In our experiment, we directly addressed these gaps. We conducted an empirical study within a realistic environment that compares three distinct levels of automation — manual driving, SAE L2, and SAE L3 driving. To estimate the cognitive load and the resources allocated for processing novel environmental stimuli, we utilized a combination of objective and subjective measurements. Finally, in interpreting our results, we consider that (automated) driving is a relatively low-demand task, at least in the present experiment. Thus, we have taken into account the aspect of top-down attentional control when interpreting our results. The top-down attention allocation has been omitted in driving studies adopting the passive oddball paradigm, presumably because all the previous studies originate in a laboratory setting and report straightforward results explainable by the competition for resources. We argue that this explanation is oversimplistic in a realistic environment.

Addressing the research gaps we identified in the current literature is crucial for designing safer and more user-centred AVs. It provides researchers and user interface designers with a better understanding of drivers' experiences, promotes realistic experiments, facilitates comparisons across automation levels, integrates objective and subjective measurements of cognitive load, and acknowledges the importance of top-down attentional control. Closing these gaps will provide valuable information for the AV design, shed light on safety concerns connected to automated driving, improve user experiences inside AVs, and shape the future of transportation.

\section{Research Questions and Hypotheses}
Our study aimed to investigate the impact of increasing vehicle automation on the cognitive processes of the driver. We focused on the attentional resources allocated to processing novel environmental stimuli and perceived cognitive load. We compared SAE L2 and L3 to manual driving using objective and subjective measures to gain an understanding of the cognitive processes involved in driver-vehicle interaction and provide insight into the neural mechanisms of drivers' behaviour. Current legal and technological constraints prevented us from an experimental design employing real traffic conditions. Therefore, we conducted an empirical test track experiment which provides high ecological validity yet a safe and controllable environment. We measured the P3a amplitude using a passive oddball task and self-reported cognitive load to understand the drivers' experience when driving manually and operating an SAE L2 and L3 AV. Our research questions and hypotheses were:

\begin{enumerate}[label=\bfseries RQ\arabic*:,leftmargin=*,labelindent=0em]
\item What are the differences in cognitive demand that drivers perceive in L2 and L3 automation compared to manual driving? 
\item Do drivers allocate attentional resources for processing environmental stimuli differently in L2 and L3 automation compared to manual driving?
\end{enumerate}

To statistically test our assumptions, we formulated the following hypotheses:

\noindent \textbf{Self-reported cognitive load:}
\begin{enumerate}[label=\bfseries H1\alph*:,leftmargin=*,labelindent=0em]
\item There is no significant difference in self-reported cognitive load (represented by the overall score of NASA-TLX) between L2 and manual driving. 
\item The self-report cognitive load (represented by the overall score of NASA-TLX) in L3 driving is lower compared to manual driving. 
\end{enumerate}

\noindent \textbf{P3a amplitude:}
\begin{enumerate}[label=\bfseries H2\alph*:,leftmargin=*,labelindent=0em]
\item There is no significant difference in frontal P3a amplitude between L2 and manual driving. 
\item The frontal P3a amplitude is decreased in manual compared to L3 driving.
\end{enumerate}

\section{Methods}
\subsection{Participants}
We recruited a total of 30 participants from the general German adult population, consisting of 16 females and 14 males, aged between 22 and 64 years (\textit{M} = 40.36 years; \textit{SD} = 13.73). Due to technical issues, data from one participant in one condition were not recorded. All participants had normal or corrected-to-normal vision and no known neurological or psychiatric diseases. They held valid German car driving licenses, with an average possession duration of \textit{M} = 19.58 years (\textit{SD} = 13.03).

Most participants were right-handed (80.00\%) and owned a car (93.33\%). None of them reported experience with automated vehicles. The participants' annual travel distances varied, with 30.00\% travelling up to 10,000 km/year, 26.67\% between 10,001 and 15,000 km/year, 23.33\% between 15,001 and 20,000 km/year, and 20.00\% travelling more than 20,000 km/year. Regarding car usage frequency, 60.00\% of participants used their car daily, 36.67\% at least once a week, and 3.33\% at least once a month.

Participants reported diverse highest obtained education, with 16.67\% completing high school, 20.00\% having vocational training, 53.33\% possessing a bachelor's or master's degree, and 3.33\% obtaining a doctoral degree. The majority of participants (73.33\%) were employed, followed by retirees (10.00\%), unemployed individuals (10.00\%), and students (6.67\%).

Participants were recruited via an external company specializing in providing subjects for research purposes. Informed consent was obtained from all participants, who received a financial compensation of 90 EUR for their involvement. The study was conducted in accordance with the Declaration of Helsinki.

\subsection{Apparatus}
\begin{figure}
\centering
\includegraphics[width=0.9\linewidth]{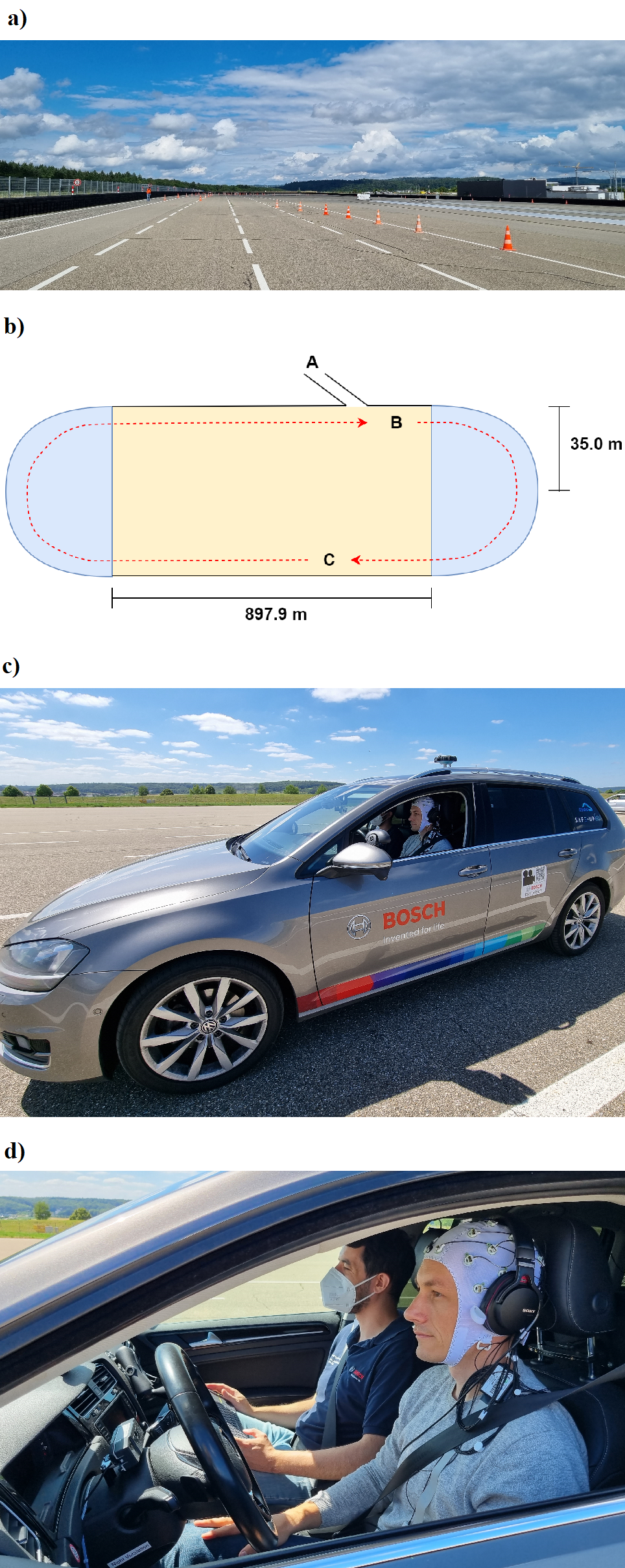}
\caption{\textbf{a)} The experiment took place on a test track in Renningen, Germany. \textbf{b)} The vehicle travelled in loops of approximately 2000 meters on a trajectory visualised by the red dotted line. In the curves (highlighted by blue), vehicle traveled with 20 km/h. In the straight stretches (highlighted by orange), vehicle traveled with 50 km/h. Point \textit{A} = entry to the test track. Point \textit{}{B} = starting and ending point of each drive. Point \textit{C} = lane departure error during L2 drive occured here. \textit{Note:} The scale is not proportional. \textbf{c)} The test vehicle was an automated vehicle prototype based on the VW Golf VII model. \textbf{d)} Participants wore a 32-channel EEG cap and headphones for the auditory stimuli presentation. A safety driver was present in the passenger seat.}
\label{Testtrack}
\end{figure}

The experiment was conducted at the Robert Bosch GmbH test track in Renningen, near Stuttgart, Germany (see Figure \ref{Testtrack}a). This test track was constructed on a landing strip of a former military airport. The vehicle traversed loops of approximately 2000 meters, as visualized in Figure \ref{Testtrack}b.

The test vehicle was a VW Golf VII with automatic transmission provided by Robert Bosch GmbH (see Figure \ref{Testtrack}c), equipped with a vehicle-in-the-loop system enabling GPS-based automated driving along a pre-defined route on test tracks. The automation function, in conjunction with the HMI, facilitated the representation of various automation levels. A safety driver, seated in the passenger seat (see Figure \ref{Testtrack}d), monitored the automation and the environment, intervening in case of imminent danger. Additionally, the safety driver was responsible for activating system errors at predetermined points on the test track.

Auditory stimuli were presented using a Fujitsu Lifebook U749 laptop and Sony MDR-1RNC headphones. The stimuli were delivered via Python 3.6 and the software package PsychoPy v2021.2.3 at 75 dB. The auditory stimuli were synchronized with the EEG signal using the Lab Streaming Layer \citep{githubGitHubSccnlabstreaminglayer}, an open-source system for the unified collection of measurement time series. Data were recorded with the LabRecorder, the default recording program of the Lab Streaming Layer.

Participants completed demographics in a paper-pencil format at the beginning of the experiment after the informed consent was signed. The perceived cognitive load was assessed using the NASA Task Load Index scale (NASA-TLX \citep{hart2006nasa}). The scale comprises six dimensions: Mental Demand, Physical Demand, Temporal Demand, Performance, Effort, and Frustration. While \textit{Mental Demand} is a key aspect of cognitive load, the other dimensions of the NASA-TLX also contribute to cognitive load by reflecting the range of demands and experiences that can impact an individual's attentional resources \citep{hart2006nasa}. Therefore, we calculate the total score of NASA-TLX by averaging across all the items to gain the overall score of cognitive load. Moreover, we administered the Karolinska Sleepiness Scale (KSS \citep{shahid2011karolinska}); however, the results are presented elsewhere \citep{figalova2023fatigue}. 

The NASA-TLX and KSS were administered as questions in UniPark, an online platform to conduct surveys. Both NASA-TLX and KSS were administered directly after each condition using Apple iPad Mini 6. 

\subsection{Experimental Conditions and Task}
Participants experienced three experimental conditions: (1)  manual driving, (2) monitoring an SAE L3 vehicle, and (3) monitoring an SAE L2 vehicle.

In the \textbf{manual condition}, participants were instructed to maintain a constant speed of 50 km/h on the straight stretches of the test track and reduce their speed to 20 km/h when navigating curves.

In the \textbf{L3 condition}, participants were informed that the vehicle could handle most situations autonomously. However, if a system boundary was reached, the vehicle would request drivers to take over control in advance. Participants were advised that at this level of automation, they could relax and enjoy the ride unless a take-over request was issued. No error or take-over request occurred during the L3 condition because our goal was to create a realistic experience of L3 driving, in which errors should not occur often.

In the \textbf{L2 condition}, participants were informed that while the vehicle would operate in an automated mode, they were required to constantly supervise the system and intervene in case of a system failure. A lane departure error was implemented, manually triggered by the safety driver approximately 7 minutes into the L2 condition. This error occurred on a straight section of the test track, causing the vehicle to drift slowly to the right. Participants needed to notice and correct the error for the vehicle to continue the ride. The error was non-critical with no imminent collision. If the participant failed to notice the error or chose not to intervene, the safety driver corrected the error by steering back in the lane, and the vehicle proceeded with the ride. In total, the error was undetected by 6 participants. In such cases, participants only experienced the correction of steering from the safety driver; no explanation or indication to intervene was given as it would disturb the sound presentation and the EEG recording. 

The automation settings for the L2 and L3 conditions were identical, with the vehicle following the same trajectory and maintaining the same velocity in both conditions. The only difference between the two conditions was the manually triggered lane departure error in the L2 condition. No secondary task was present in any of the conditions. The order of the conditions was counterbalanced among participants using a balanced Latin square design.

\subsection{Stimuli of the Passive Oddball Task}
During the three experimental conditions, participants were subjected to a passive oddball task, which consisted of three types of task-irrelevant auditory stimuli:

\begin{enumerate}
\item[(1)]\textbf{Frequent distractors,} with 450 presentations in each condition (probability of presentation: 71.43\%);
\item[(2)]\textbf{infrequent distractors,} with 90 presentations in each condition (probability of presentation: 14.29\%);
\item[(3)]\textbf{novel environmental distractors,} with 90 presentations in each condition (probability of presentation: 14.29\%).
\end{enumerate}

The frequent and infrequent distractors consisted of two easily distinguishable beep tones (pure 700 Hz and 300 Hz tones); their presentation probability was counterbalanced across participants. For the environmental distractors, 30 audibly distinct complex sounds (e.g., human laughter, helicopter, animal sounds) were selected from a database provided by the New York State Psychiatric Institute \citep{fabiani1996naming}. We used the same set of stimuli as \citet{scheer2016steering} and \citet{scheer2018auditory}. These environmental distractors did not include sounds with strong emotional valence (e.g., sexual sounds). All stimuli featured a 10 ms rise and fall, with a total duration of 336 ms for each stimulus. The inter-stimulus interval was randomized, ranging from 1300 to 1700 ms. Environmental sounds were presented in a quasi-random order without repetition and always followed at least one frequent distractor. 

Each of the three experimental conditions lasted approximately 17 minutes. In every condition, participants experienced three consecutive blocks of sound presentation. Each block consisted of 210 auditory stimuli (150 frequent, 30 infrequent, and 30 novel environmental distractors). Each block lasted approximately five minutes and 30 seconds, during which auditory stimuli were presented. The three blocks were separated by 20-second intervals, during which participants continued the driving task without sound presentations. The three blocks of auditory stimuli were identical; therefore, each of the 30 distinct novel environmental sounds was played once in each block, resulting in three repetitions of each stimulus in each condition.  

The auditory stimuli were presented in each of the three experimental conditions. We informed participants that a presentation of sounds would begin once they started driving. They were instructed that these sounds were not important for their task and they could ignore them. 

\subsection{Procedure}
Upon arrival, participants signed the informed consent form and completed a demographic questionnaire. Following this, the EEG headset was set up, and participants were driven to the test track. Once seated in the driver's seat, they received safety training from the safety driver. Next, they experienced a practice drive. In this practice drive, participants manually drove one loop around the test track, with the safety driver guiding the desired trajectory and velocity of the vehicle. This drive was not recorded and served solely as training. Subsequently, participants watched an introductory video containing general information about the experiment and its procedure. The three driving conditions (manual, L2, and L3 driving) followed in a counterbalanced order.

At the beginning of each of the three driving conditions, participants viewed an introduction video. This video provided information about the current level of automation of the vehicle (no automation, L2, or L3 automation). The video outlined the system's capabilities and the driver's tasks during the drive. After watching the video, participants could ask further questions. Once they understood their task, they were asked to proceed with the task. Task-irrelevant auditory probes were played as soon as the vehicle began moving.

At the end of each of the three conditions, participants manually drove back to the starting point of the test track and filled in the NASA-TLX and KSS. Therefore, we obtained the self-reported data at three measuring points, immediately at the end of each condition. After completing the final condition, participants were driven back to the facility. After they washed their hair, a short debriefing session was provided. The experiment's objectives were explained, and participants could ask questions and provide feedback about the experiment. The entire experiment lasted approximately 120-150 minutes, depending on the individual differences in the EEG set-up time.

\subsection{EEG Signal Acquisition and Processing}
The EEG was recorded using 32 channels placed according to the international 10-20 system. We utilized the ActiCAP set (Brain Products GmbH, Germany) with active shielded electrodes and a LiveAmp amplifier. The impedance of each electrode was maintained below $25 k\Omega$ using SuperVisc electrolyte gel (Easycap GmbH, Germany). The data were online referenced to the FCz electrode and digitized with a 1000 Hz sampling rate using the Lab Streaming Layer framework \citep{kothe2012lab}. The data were preprocessed in Matlab version R2022a, employing EEGlab v2022.1 with the ERPlab v9.00 \citep{delorme2004eeglab}, IClabel v1.4 \citep{pion2019iclabel}, bemobil-pipeline 1.9 \citep{klug2022bemobil}, clean\_rawdata v2.7, dipfit v4.3 \citep{delorme2011eeglab}, xdfimport v1.18 \citep{kothe2012xdfimport}, and zapline-plus v1.2.1 \citep{klug2022zapline, de2020zapline} plugins. 

Data preprocessing followed the BeMoBil pipeline \citep{klug2022bemobil}. We followed the recommendations that are provided by \citet{klug2022bemobil} and are available in the Matlab documentation. All of the steps that required setting parameters and/or thresholds are documented in this section. The continuous EEG data were downsampled to 500 Hz and band-pass filtered between 0.1 and 100 Hz. Power line artefacts were removed using the ZapLine Plus plugin \citep{de2020zapline, klug2022zapline}. Channels with less than $\textit{r} = .78$ correlation to their robust estimate in over 50\% of the time were interpolated (on average, \textit{M} = 5.13 channels, \textit{SD} = 1.86). All channels were offline re-referenced to an average reference. The data were high-pass filtered with a cut-off frequency of 1.25 Hz to perform adaptive mixture independent component analysis (AMICA). Automatic rejection of bad data portions was enabled, and AMICA was conducted with ten iterations. The spatial filter was then copied into the original dataset, and components were classified using IClabel. Components most likely originating from brain activity were retained.

After cleaning the raw data using the copied AMICA weights, we applied a second-order Butterworth filter and band-pass filtered the data between 0.1 and 30 Hz. Next, continuous artefact rejection was performed with a frequency range for thresholding between 20 and 30 Hz. The upper-frequency threshold was set to 10 dB with a 0.5 s epoch length. A minimum of four contiguous epochs was required to label the region as artifactual, with an additional 0.25 s neighbouring signals on each side.

Finally, we extracted epochs from -200 to 800 ms relative to the stimulus onset. Baseline correction was applied, and epochs containing step-like activity were flagged as artifactual. We utilized a threshold of 50 $\mu$V in a 200 ms moving window width with a 10 ms moving window step. On average, 286.29 epochs (\textit{SD} = 238.27) were rejected. For the final analysis, each dataset contained, on average, 1423.14 epochs (\textit{SD} = 262.84) across the three conditions. Consequently, 84.15\% of the total epochs (\textit{SD} = 13.24\%) were included in the data analysis.

\subsection{ERP Parametrization and Statistical Analysis}
The preprocessed data were averaged for each channel and experimental condition. A difference wave (DW) was computed as the difference between the ERP elicited by the environmental distractor and the ERP elicited by the frequent distractor. We decided to compute DWs because, in a raw ERP waveform, different cognitive processes are superimposed on one another, and the P3a response might be overlaid with other neural activity that is not specifically related to the detection of the auditory stimulus. Subtracting the ERP response to standard stimuli from the ERP response to oddball stimuli allowed us to isolate the specific change in neural activity associated with detecting the oddball \citep{luck_2014}. Other researchers using the same paradigm have chosen the same approach (e.g., \citet{scheer2016steering, van2021effect, wester2008event, kramer1995assessment}). The amplitudes of the frequent and infrequent beep tones are not reported, as the auditory P3a is generated by unexpected novel sounds. 

For the statistical analysis, we focused on the frontal region. We analysed the frontal patch consisting of the Fz, FC1, and FC2 electrodes. We chose the frontal area as the P3a reflects frontal lobe activity related to the perception of novel distractors \citep{polich2007updating}. Moreover, it has been shown that frontal P3a exhibit the largest P3a amplitudes \citep{scheer2016steering}. We used the patch of three neighbouring frontocentral electrodes to improve the signal-to-noise ratio.

Statistical analysis was conducted using JASP 0.16.4, with the general level of significance set to 0.05. For self-reported result comparison, we employed repeated measures ANOVA. When sphericity was violated, we applied the Greenhouse-Geisser correction (when $\epsilon < 0.75$) or the Huynh-Feldt correction (when $\epsilon > 0.75$). Holm post-hoc tests were utilized for pairwise comparisons. Effect sizes are reported using $\omega_{}^{2}$ (for rmANOVAs) and Cohen's d (for post-hoc tests).

We used linear mixed-effects models to compare EEG results with the maximum likelihood method for model fitting. Linear mixed-effects models are a more robust choice in case of missing data than traditional ANOVA models \citep{bagiella2000mixed}. Due to missing data points, linear mixed-effects models were chosen as a more appropriate alternative to repeated-measures ANOVA in our design.

We employed the collapsed localizer method to quantify ERP component amplitudes. This method provides an objective, data-driven way of identifying the time window when an ERP component is most prominent and reduces the potential for bias or error in the selection of the parameters \citep{luck_2014, luck2017get}. A grand average (GA) peak for the three conditions at the frontal patch was determined at 354 ms post-stimulus. The mean amplitude was calculated between 329 and 374 ms post-stimulus (GA peak +/- 25 ms). The model was specified as:

\begin{equation} \label{LMM}
Amplitude \sim Level + (1\mid Subject)
\end{equation}

where \textit{Amplitude} is the mean P3a amplitude in the selected time window, \textit{Level} is the fixed effect of the level of automation, and \textit{Subject} is the random effect of each test subject. To assess contrasts between mean amplitudes, we computed estimated marginal means with Holm p-value adjustment for differences between L2 and manual driving and between L3 and manual driving.

\section{Results}
\subsection{Self-reported Cognitive Load}
A repeated measures ANOVA with a Huynh-Feldt correction determined that the self-reported cognitive load, quantified as the mean NASA-TLX score, differed between the three conditions (\textit{F}(1.683, 48.807) = 10.097, \textit{p} \textless .001, $\omega_{}^{2}$ = 0.108). Post hoc testing using the Holm correction revealed that L2 driving was perceived as more demanding than L3 driving (mean difference = 2.88, \textit{p} \textless .001, \textit{d} = 0.834), and manual driving was perceived as more demanding than L3 driving (mean difference = 3.508, \textit{p} = .002, \textit{d} = 0.699). There was no difference between L2 and manual driving (mean difference = 0.467, \textit{p} = .501).  In accordance with our hypotheses H1a and H1b, we observed no difference in self-reported cognitive load between manual and L2 driving, and the self-reported cognitive load was significantly lower in L3 driving than in manual driving.  

\begin{figure*}
\centering
  \includegraphics[width=1\linewidth]{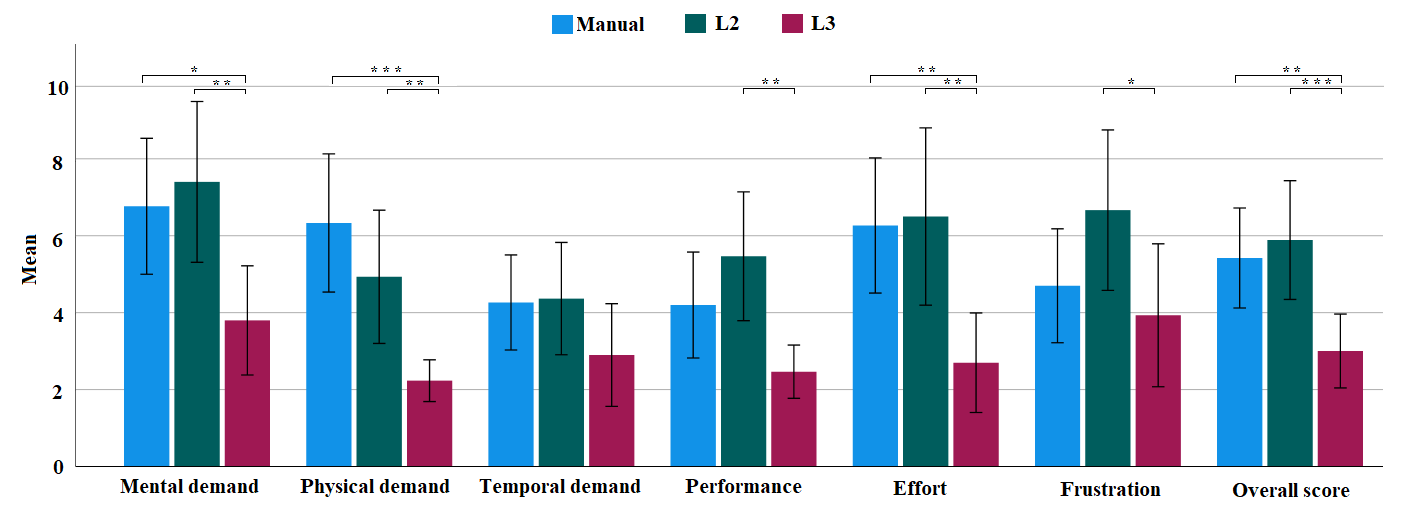}
  \caption{The comparison of the mean score of the six dimensions of NASA-TLX and the overall score. Significant differences are marked with an asterisk (* \textit{p} \textless .05, ** \textit{p} \textless .01, *** \textit{p} \textless .001). The error bars represent 95\% confidence intervals.}
  \label{Results_NASA}
\end{figure*}

To further understand the factors that influence the perceived cognitive load in different levels of automation, we analysed the six dimensions of NASA-TLX. Figure \ref{Results_NASA} visualizes the mean score of each dimension and the overall NASA-TLX score, and the results of multiple comparisons between the three conditions using post-hoc tests with the Holm correction. 

The mean scores of \textbf{mental demand} (e.g. thinking, deciding, calculating, remembering, looking, searching) differed significantly between the levels of automation (\textit{F}(2, 58) = 7.159, \textit{p} = .002, $\omega_{}^{2}$ = 0.085). Using Holm post-hoc test revealed a significant difference between manual (\textit{M} = 6.767, \textit{SD} = 4.732) and L3 driving (\textit{M} = 3.800, \textit{SD} = 3.800, mean difference = 2.967, \textit{p} = .010, \textit{d} = 0.622); and between L2 (\textit{M} = 7.400, \textit{SD} = 5.599) and L3 driving (mean difference = 3.600, \textit{p} = .002, \textit{d} = 0.755). There was no difference between manual and L2 driving (mean difference = 0.633, \textit{p} = .535, \textit{d} = 0.133). These results suggest that mental demand was lower in L3 driving than in manual and L2 driving, while no difference was observed between manual and L2 driving. 

The mean scores of \textbf{physical demand} (e.g. pushing, pulling, turning, controlling, activating) differed significantly between the levels of automation (\textit{F}(2, 58) = 11.881, \textit{p} \textless .001, $\omega_{}^{2}$ = 0.145). Using Holm post-hoc test revealed a significant difference between manual (\textit{M} = 6.333, \textit{SD} = 4.809) and L3 driving (\textit{M} = 2.233, \textit{SD} = 4.638, mean difference = 4.100, \textit{p} \textless .001, \textit{d} = 1.039); and between L2 (\textit{M} = 4.933, \textit{SD} = 4.638) and L3 driving (mean difference = 2.700, \textit{p} = .005, \textit{d} = 684). There was no difference between manual and L2 driving (mean difference = 1.400, \textit{p} = .107, \textit{d} = 0.355). These results suggest that L3 driving requires less physical effort than manual and L2 driving. No difference was found between manual and L2 driving in terms of physical demand.

The mean scores of \textbf{temporal demand} (how much time pressure participants felt) did not differ significantly between the manual (\textit{M} = 4.267, \textit{SD} = 3.311), L2 (\textit{M} = 4.367, \textit{SD} = 3.908), and L3 (\textit{M} = 2.900, \textit{SD} = 3.575) driving (\textit{F}(2, 58) = 2.732, \textit{p} = .073, $\omega_{}^{2}$ = 0.021). 

The mean scores of \textbf{performance} (how satisfied were participants with their performance) differed significantly between the levels of automation (\textit{F}(2, 58) = 6.782, \textit{p} = .002, $\omega_{}^{2}$ = 0.094). Using Holm post-hoc test revealed a significant difference between L2 (\textit{M} = 5.467, \textit{SD} = 4.485) and L3 driving (\textit{M} = 2.467, \textit{SD} = 1.852, mean difference = 3.000, \textit{p} = .002, \textit{d} = .852). There was no difference between manual (\textit{M} = 4.200, \textit{SD} = 3.690) and L2 driving (mean difference = 1.267, \textit{p} = .127, \textit{d} = 0.360). There was no difference between manual and L3 driving (mean difference = 1.733, \textit{p} = .077, \textit{d} = 0.492). The score on this subscale is reversed. Therefore, the results suggest that participants' satisfaction with their performance was higher in L3 compared to L2 driving. No differences were found between manual and L2 driving or manual and L3 driving in terms of satisfaction with one's performance.

The mean scores of \textbf{effort} (how hard participants had to work to accomplish their performance) differed significantly between the levels of automation (\textit{F}(1.565, 45.388) = 7.699, \textit{p} = .003, $\omega_{}^{2}$ = 0.098). Using Holm post-hoc test revealed a significant difference between manual (\textit{M} = 6.267, \textit{SD} = 4.705) and L3 driving (\textit{M} = 2.700, \textit{SD} = 3.466, mean difference = 3.567, \textit{p} = .003, \textit{d} = 0.727); and between L2 (\textit{M} = 6.500, \textit{SD} = 6.174) and L3 driving (mean difference = 3.800, \textit{p} = .003, \textit{d} = .774). There was no difference between manual and L2 driving (mean difference = 0.233, \textit{p} = .831, \textit{d} = 0.048). These results suggest that L3 driving requires less effort than manual and L2 driving. No difference was found between manual and L2 driving regarding perceived effort.

The mean scores of \textbf{frustration} (whether participants felt insecure, discouraged, irritated, stressed and annoyed during their task) differed significantly between the levels of automation (\textit{F}(2, 58) = 3.669, \textit{p} = .0032, $\omega_{}^{2}$ = 0.039). Using Holm post-hoc test revealed a significant difference between L2 (\textit{M} = 6.667, \textit{SD} = 5.585) and L3 driving (\textit{M} = 3.933, \textit{SD} = 4.975, mean difference = 2.733, \textit{p} = .033, \textit{d} = .559). There was no difference between manual (\textit{M} = 4.700, \textit{SD} = 3.967) and L2 driving (mean difference = 1.967, \textit{p} = .128, \textit{d} = 0.402). There was no difference between manual and L3 driving (mean difference = 0.767, \textit{p} = .464, \textit{d} = 0.157). These results suggest that participants felt more frustrated in L2 than L3 driving. However, there were no significant differences between manual and L2 driving or manual and L3 driving in terms of frustration levels.

\textbf{Overall}, mental and physical demands were significantly lower for L3 driving compared to manual and L2 driving. No differences were found in temporal demands among the three levels. Drivers evaluated their performance to be worse during L2 driving than L3 driving, while the effort was lower for L3 driving than for manual and L2 driving. Frustration was lower in L3 driving than L2 driving, but no significant differences were found between manual and the other two levels.

\subsection{P3a Amplitude}
\begin{figure*}
\centering
  \includegraphics[width=0.9\linewidth]{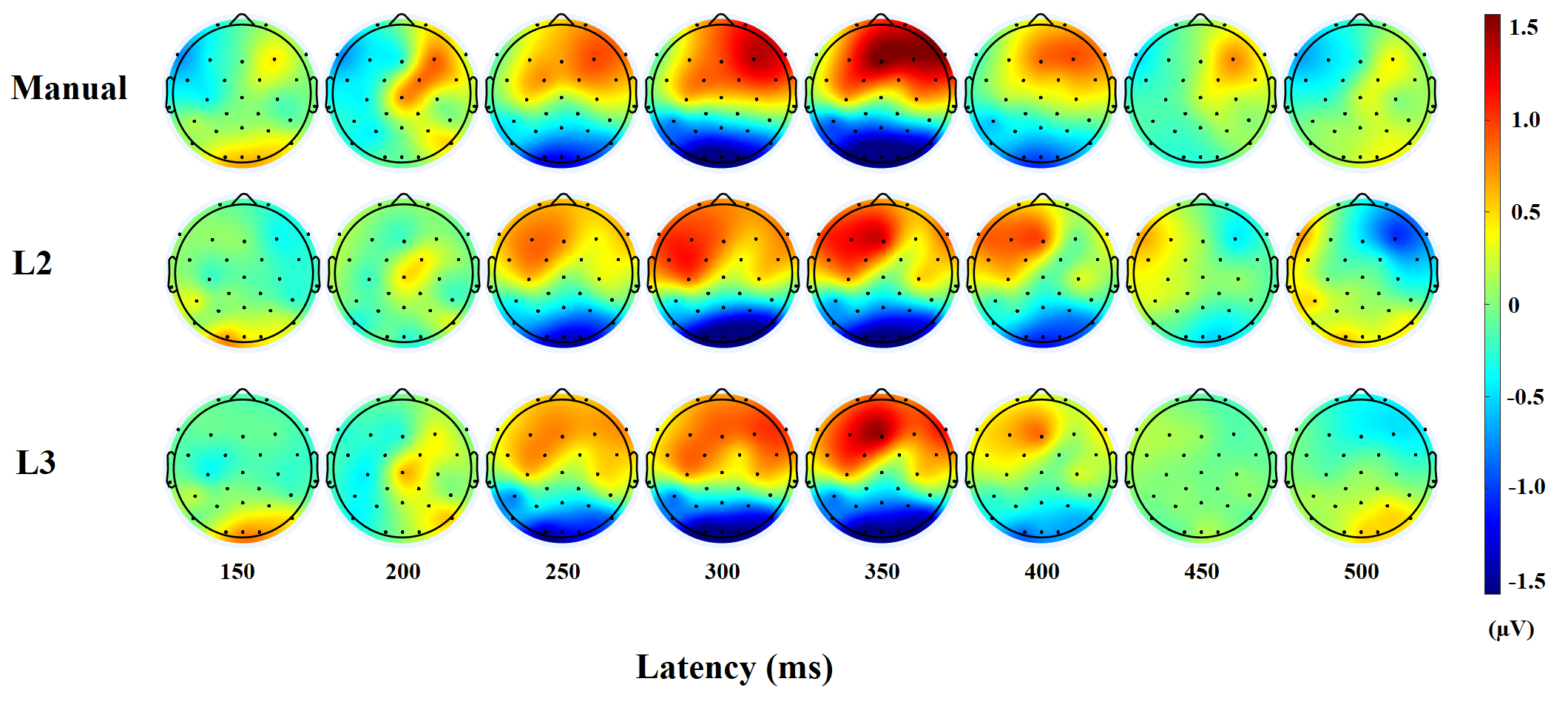}
  \caption{Topographical map for various 50-ms intervals from 150 ms after stimulus onset to 500 ms after stimulus onset. The map displays the difference between environmental and frequent auditory cues.}
  \label{Results_topoplot}
\end{figure*}

Figure \ref{Results_topoplot} provides a visualization of the scalp's electrical activity distribution in response to stimulus presentation. This topographical map highlights the variance in neural activity achieved by subtracting the response to frequent stimuli from the response to novel environmental stimuli. Furthermore, subtracting the ERP responses improves the signal-to-noise ratio and removes the superimposed neural activity, which is not related to stimulus processing. The plot illustrates the relative disparities in the activation triggered by novel environmental stimuli across different conditions. 

The figure visualises the information-processing cascade associated with stimulus detection, categorization, and evaluation, described in the three-stage involuntary attention-switching framework. In the perception stage, identified around 100-150 ms, the brain recognises the presence of the stimulus. However, at this stage, the difference between frequently encountered and novel stimuli is negligible because the brain is primarily involved in capturing the perceptual essence of the stimuli, regardless of their familiarity. Subsequently, in the categorisation stage, evident around 250-400 ms after the introduction of stimuli, the brain engages in a cognitive evaluation of these stimuli. At this point, the contrast between frequent and novel stimuli becomes pronounced. The novel environmental sounds necessitate a greater allocation of processing resources as compared to the frequently encountered stimuli. Once the distracting stimuli are categorised as irrelevant, an attention switch occurs, and the difference between the frequent and novel stimuli is minimal again.  

\begin{figure*}
\centering
  \includegraphics[width=0.9\linewidth]{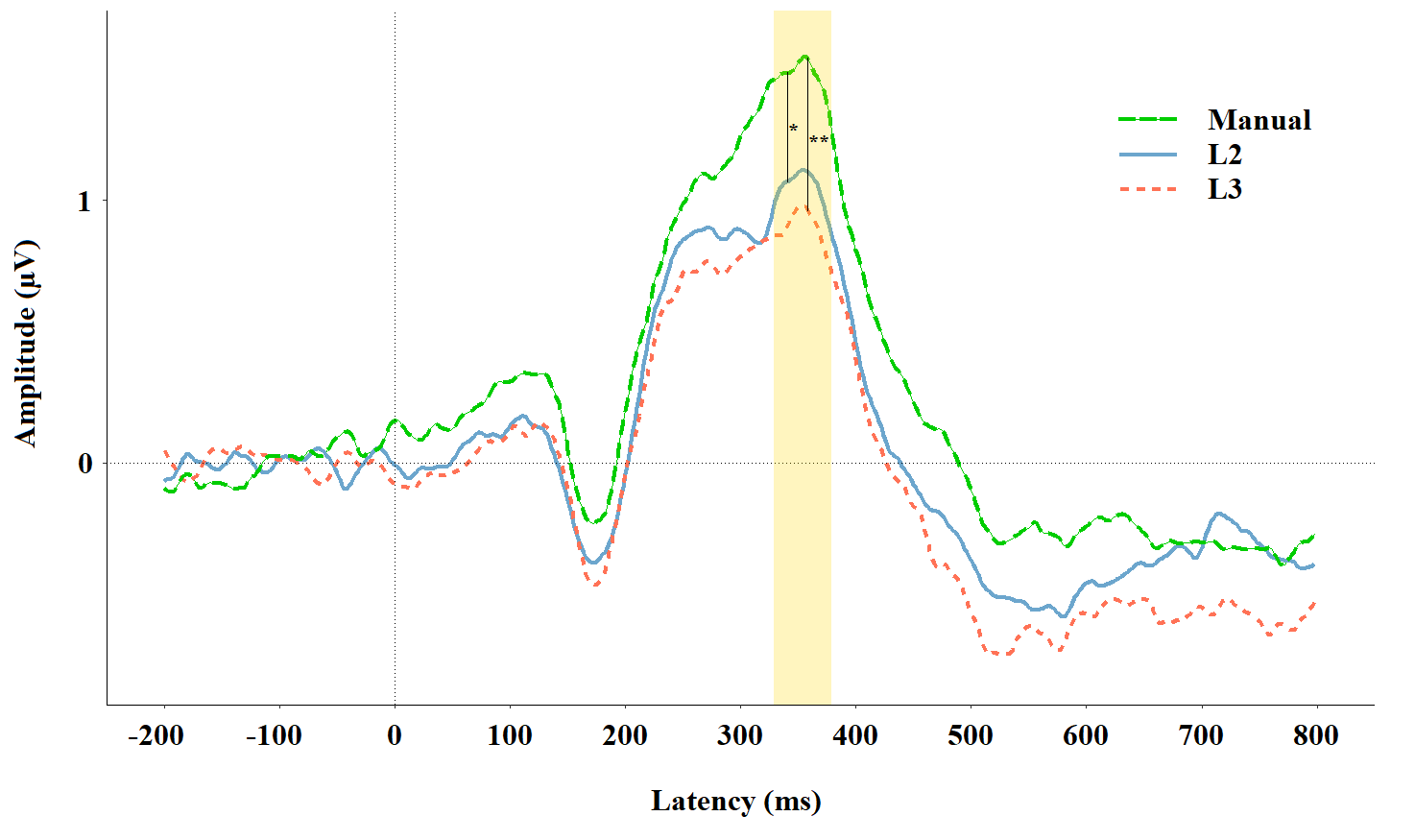}
  \caption{Difference wave (environmental-frequent) ERPs in the three experimental conditions. Mean amplitudes in the yellow-highlighted area (329 and 379 ms) were submitted for statistical comparisons (* \textit{p} \textless .05, ** \textit{p} \textless .01).}
  \label{Results_ERP}
\end{figure*}

Figure \ref{Results_ERP} presents the difference waves in the time domain. The mean amplitude between 329 and 379 ms on the frontal patch (Fz, FC1, FC2) was \textit{M} = 1.436 \si{\micro\volt} (\textit{SD} = 1.243) for manual driving, \textit{M} = 0.920 \si{\micro\volt} (\textit{SD} = 1.006) for L3 driving, and \textit{M} = 1.046 \si{\micro\volt} (\textit{SD} = 0.921) for L2 driving. Utilizing a linear mixed model (BIC = 269.584, deviance = 247.140), we observed a significant main effect of the level of automation (beta = 0.333, \textit{t} = 2.757, \textit{p} = .008). We identified a difference between L3 and manual driving (\textit{M} = 0.576, \textit{SE} = 0.208, \textit{p} = .011). Additionally, we discovered a difference between L2 and manual driving (\textit{M} = 0.422, \textit{SE} = 0.211, \textit{p} = .045).

Our results do not support the H2a and H2b hypotheses. We anticipated no significant difference between the mean P3a amplitude in manual and L2 driving, suggesting a similar allocation of attentional resources for processing environmental sounds. However, we found a difference between manual and L2 driving. Furthermore, we assumed that the P3a amplitude would increase in L3 compared to manual driving, implying that more attentional resources are devoted to processing environmental sounds. Contrary to our assumption, we observed a significant decrease in amplitude in L3 compared to manual driving. In summary, our results indicate that the greatest amount of attentional resources for processing environmental sounds was utilized during manual driving and reduced in both L2 and L3 driving.

\section{Discussion}
This study investigated the cognitive mechanisms underlying driver-vehicle interactions. Our goal was to get a better understanding of the variations in self-reported cognitive load and attentional resource allocation among manual driving and SAE L2 and SAE L3 automated driving. We conducted an empirical experiment with 30 drivers who operated a prototype automated vehicle on a test track. Throughout the different automation levels, drivers experienced a passive oddball task and encountered three categories of task-irrelevant auditory cues: frequent beeps, infrequent beeps, and infrequent novel environmental sounds. We analysed the amplitude of the P3a ERP elicited by the environmental sounds, which served as an indicator of attentional resources employed in processing these sounds. Additionally, we assessed participants' self-reported cognitive load after each driving session. In this section, we discuss our findings concerning self-reported subjective cognitive load, the P3a amplitude as an objective indicator of attentional resource allocation, the differences between the objective and subjective measurements, discrepancies with previous studies, implications of our findings, and the limitations and future research directions.

\subsection{Subjective Cognitive Load}
To assess the perceived subjective cognitive load, we analysed the total score of NASA-TLX obtained by averaging across all items. The results show that drivers perceived L3 driving as the least demanding, while no significant difference was observed between manual and L2 driving regarding self-reported cognitive load. Upon examining the six factors of NASA-TLX (mental, physical, and temporal demand, performance, effort, frustration), L3 driving required lower mental and physical demand and less effort than manual and L2 driving. Interestingly, there was no difference in self-reported physical demand between manual and L2 driving, even though the actual physical demand was higher in manual driving but comparable for L2 and L3 driving.

Drivers rated their performance more favourably in L3 than in L2 driving and experienced less frustration with the L3 system. The performance factor asked participants how successful they were in performing the task and how satisfied they were with their performance. The frustration factor asked how irritated, stressed, and annoyed versus content, relaxed, and complacent drivers felt during the task. We interpret these results in a way suggesting that drivers perceived overall lower demands in L3 driving. Manual and L2 driving were perceived as somehow similar with regard to cognitive load.

These findings are consistent with prior research of \citet{stapel2017driver, stapel2019automated}. The authors compared objective and subjective mental workload during manual and L2 driving in real traffic on Dutch highways. Their findings suggest that automation-inexperienced drivers perceived similar subjective workloads in manual and L2 driving. When assessing our data, we observed a similar trend.

However, the findings presented in this study are inconsistent with the findings of \citet{kim2023partially} and \citet{varhelyi2021driving}. \citet{varhelyi2021driving} compared manual and L3 driving on a German highway. The results showed no differences with regard to perceived mental workload between the manual and L3 conditions. This is different from the present study, as we observed a significant decrease in perceived cognitive load during L3 driving. We assume that this difference is due to the different environmental settings of the experiments. The discrepancies between a simulator, test track, and real traffic are further discussed in section \ref{simvsreal}. 

Moreover, \citet{kim2023partially} conducted an experiment with L2 vehicles in real traffic in the United Kingdom. They reported an increase in perceived mental workload connected to the use of automation. The authors interpret these findings through several factors. Primarily, they point to the high mental demand required to monitor the automated vehicle. However, they also note that drivers were required to develop a new mental model to comprehend the automation status. Further, the study identified high levels of drivers' frustration in the automation mode and the increased risk perception associated with complex real-world conditions. The authors suggest that the increased mental workload may be drivers' natural response to the tangible risks of real-world driving, potentially leading to a greater cognitive load than that experienced in simulated environments. It is important, however, to mention the small sample size of the experiment (N = 8) and the fact that all participants were members of the project team. 

Another discussion point arises from the use of the overall score of NASA-TLX. Even though calculating the perceived cognitive load by averaging across the six dimensions of NASA-TLX is a common practice in the field, some authors suggest that this approach is not mathematically meaningful \citep{bolton2023mathematical}. Therefore, we provide both overall and per-dimension analysis of cognitive load to provide a possible comparison with other studies.

Overall, we argue that our results regarding perceived cognitive load are complimentary with the previous findings. We observed similar trends as \citet{stapel2017driver, stapel2019automated}. Even though the results are inconsistent with the results of \citet{kim2023partially} and \citet{varhelyi2021driving}, we believe that the different environmental settings is the primary factor in the observed differences. 

\subsection{P3a Amplitude and Attentional Resource Allocation}
The amplitude of the P3a ERP component can be considered an objective measure of the attentional resources allocated towards processing environmental auditory stimuli and can serve as a measure of cognitive load \citep{kramer1995assessment, polich2007updating}. Some studies also suggest that the neuroelectric response to novel auditory stimuli can be modulated by top-down attention allocation \citep{cahn2009meditation}. In light of the literature, our initial hypotheses were that there would be no observable difference between the P3a amplitude in manual and L2 driving and that the P3a amplitude would be increased during L3 driving. However, our observations diverged from these expectations, revealing the highest P3a amplitude during manual driving. 

This outcome deviates from what has been reported in several other studies \citep{wester2008event, scheer2016steering, van2018susceptibility, van2021effect}. These studies report the highest P3a amplitude in stationary conditions where the vehicle was not moving, with a reduction in automated driving relative to stationary conditions and a further decrease in manual driving. The authors proposed that cognitive load was highest when driving manually, with automated driving demanding less cognitive resources, and the least cognitive load present during the stationary condition.

In our view, however, interpreting these results purely through the lens of cognitive resource competition might be an oversimplification. This interpretation is predicated on the idea of limited, finite processing resources, with information competing for this limited pool. It is important to note, however, that all of the studies mentioned earlier deployed an experimental paradigm requiring relatively low cognitive effort. This brings into question whether a \textit{floor effect} could be observed when the measure loses its sensitivity. We suggest that other factors could be contributing to the modulation of the P3a amplitudes.

In the stationary condition, drivers were not performing a task. When left in such situations for a longer period of time, they likely felt bored. Boredom could explain why they allocated more resources to the processing of the environmental sounds, as they could serve as a distraction from the boring situation. 

Moreover, the P3a amplitude was decreased in manual compared to automated driving in the above-mentioned studies. However, the simulated environment could explain this observation. As the simulator is unable to fully reproduce a realistic driving experience \citep{groh2019simulation, hock2018design, wynne2019systematic, carsten2011driving}, the findings might have limited relevance in real-world settings. This matter is further discussed in the section \ref{simvsreal}. 

Additionally, other studies have not considered the change in drivers' roles when interpreting the results. The changed role could imply that the environmental sounds, which could be rather irrelevant for the drivers of automated vehicles, become more relevant when driving manually. The change in relevance could be because manual driving requires integrating different perceptual information to create a unified perception of the world \citep{spence2020crossmodal}. Environmental sounds are potentially important in manual driving, as drivers must constantly update their perceptual-motor loop to remain calibrated to the vehicle and their steering task \citep{mole2019getting}. With increasing automation, drivers are expected to focus their attention more inside the vehicle due to the growing use of tactile displays, alerts, and warning signals inside vehicles \citep{spence2020crossmodal}; hence, the importance of environmental information decreases. Therefore, we argue that cognitive goals could enhance the bottom-up processing of environmental stimuli in manual driving. This effect could be attenuated in semi-automated driving. However, further investigation on this topic is necessary to draw conclusions about the change in spatial attention allocation. 

Finally, previous experience with automation can influence drivers' experience \citep{lohani2021no, stapel2019automated}. All the participants in the present study were active drivers but had no previous experience with automation. Therefore, manual driving was an easy task for them, while automation was unusual. A novelty effect may influence our results; drivers could have allocated less attentional resources to process the environmental stimuli due to the excitement of experiencing automated driving for the first time.

Overall, the changes in P3a amplitude in the present study were different than we expected based on the literature review. We argue that the discrepancy originates mainly in the different environmental settings of the experiments (simulator in the previous studies vs real vehicle in our study), the changing role of drivers in automated vehicles, and the novelty effect due to our sample characteristics (automation-inexperienced drivers). We explain this observation through the lens of top-down attention control, which was omitted in the previous studies. 

\subsection{Discrepancies between Simulators and Realistic Environments} \label{simvsreal}
One of the major points for discussion is the effect of environmental settings on the results of studies in the field of driving research. Many researchers report discrepancies in a simulator and real-world findings in regards to the speed of driving \citep{zoller2019driver}, lateral position \citep{daurat2013lorazepam}, eye fixations \citep{fors2013simulator}, heart rate \citep{johnson2011physiological}, or mental workload \citep{mueller2015driving} (for a review, see \citet{wynne2019systematic}). 

Although technological progress allows conducting experiments in highly immersive simulators, the realism of such experience remains limited as simulators can not completely replicate the real world. Participants of simulator experiments remain aware of the safe nature of the simulator environment, which modulates their risk perception, and subsequently impacts their behaviour \citep{carsten2011driving}. Moreover, the technological properties of the simulator modulate the driving experience \citep{figalova2022system}. \citet{sadeghian2018feel} suggests that real-world conditions for simulation, e.g., motion, can clarify mixed findings on driver-vehicle interactions in automated driving. 

Although test tracks cannot replace real traffic conditions due to the safe environment and reduced complexity, it allows researchers to study driving in a more realistic setting. The results of this study suggest that the P3a findings obtained on a test track differ from those in a driving simulator. We argue that the environment has a major impact on the phenomena studied in driving research. \citet{chuang2017using} previously came to similar conclusions using EEG in a laboratory and a driving simulator to understand the differences in drivers' behaviour. Subsequently, we recommend a critical approach to findings originating in a driving simulator. Moreover, we stress the importance of real-world testing. 

\subsection{Discrepancies between Objective and Subjective Measures}

The discrepancy between objective and subjective measures has been previously described in the literature, for example, by \citet{stapel2019automated, stapel2017driver} or \citet{large2017exploring}. These discrepancies, however, do not necessarily mean contradicting results. \citet{de1996measurement} suggests that different approaches to measuring cognitive load are sensitive to different components of cognitive load and recommends combining measures from different categories (self-report, performance-based, and physiological measures) to gain a comprehensive understanding of cognitive load.

The current literature jointly reports the modulations of P3a amplitude to be an effect of cognitive load. However, these studies were generally conducted in laboratory settings and targeted very specific and well-defined problems. However, real-world driving is a complex task in a complex environment. Therefore, interpreting the P3a amplitude only through the lens of resource competition could be misleading. As the results of \citet{cahn2009meditation} suggest, top-down attention control could be another factor modulating the measure. We argue that other factors, such as novelty or boredom, could be of importance as well. 

The results of this experiment suggest that cognitive load might not be the only factor influencing the P3a amplitude, which is currently the explanation provided by the majority of literature. Therefore, we believe that using P3a amplitude as a direct measurement of cognitive load might omit other important factors. We propose that top-down attention control is another important factor. Further studies must be carried out to specify what exactly is reflected in the P3a amplitude.

\subsection{Implications}
Automation promises to increase traffic safety, efficiency, and the convenience of travelling. Until fully autonomous vehicles are available, humans remain responsible for the operation of semi-autonomous systems and, therefore, remain a crucial safety factor \citep{stanton2009human, bucsuhazy2020human}. The use of neuroscientific measures in driving research can help inform the design of user interfaces that improve how drivers process the conveyed information \citep{glatz2018use}, subsequently improving safety and user experience of automated driving. The results presented in this paper have several implications for human-computer interaction research and user interface design for future automated vehicles. 

Our results suggest that drivers of L2 and L3 vehicles allocate fewer attentional resources to process environmental information than when driving manually, potentially hindering drivers' situational awareness. These findings stress the need to design interfaces that efficiently convey information to drivers who might be distracted or out of the loop. 

Level 2 driving was perceived to be comparably demanding as manual driving. This suggests that potential benefits of L2 systems in terms of increased driving comfort may not be easily reaped - presumably especially when drivers are novices or driving durations are shorter. Moreover, the effects of long-term exposure to automated driving should be addressed in future studies, because these implications might differ for experienced drivers. 

Furthermore, the present results highlight the importance of driver monitoring systems that can assess drivers' states in real time. Understanding drivers' state can inform intelligent user interfaces, which should dynamically adapt to drivers' current needs. These interfaces should manage drivers' attention and task load, ensuring optimal user interaction and engagement, subsequently improving the user experience.

The low cognitive demand experienced by drivers in higher levels of automation could negatively impact traffic safety. As drivers might be distracted, out of the loop, and not calibrated to the vehicle dynamics before the onset of a take-over request, regaining full manual control could lead to safety-critical situations. Using collaboration between the driver and the vehicle allows to combine the advantages of high sensor precision with superior human decision-making. Such cooperation should be adaptive and based on the input from driver monitoring systems in order to guide drivers successfully and provide relevant, efficient and accepted support. The collaborative approach could, therefore, improve user experience and safety of automated driving.

The results of the present study point out the importance of real-world testing. The human-computer interactions should be addressed and evaluated in a naturalistic environment, even though the experimental control might be decreased. Natural interactions provide invaluable insights necessary for a comprehensive understanding of human-computer interactions. 

\subsection{Contribution}
This study contributes to a better understanding of cognitive mechanisms that underlay driver-vehicle interactions. Our results bring new evidence about the self-reported cognitive load and attentional resource allocation among manual, SAE L2, and SAE L3 driving in a realistic environment. To our best knowledge, comparing three levels of automation using both objective and subjective measures in a realistic environment has not been previously published. Our findings highlight the need for further investigation of the factors influencing the allocation of attentional resources during driving, including the role of automation experience, experimental environment, and individual differences. Moreover, our results underscore the importance of considering both bottom-up and top-down attention control in P3a experiments which use tasks inducing low to medium levels of cognitive load. Future experiments can address this issue methodologically and study the effect of various factors on the P3a amplitude. 

Our results highlight several points that may be considered for improving the design of automated driving systems, especially for L2 systems:  1. Drivers may not readily experience the benefits of partial automation 2. Drivers' situational awareness may be diminished when using the automated system. These points may be considered by non-technical means, such as management of expectations through appropriate advertisement or drivers' training and by technical means, such as the design of driver-vehicle interfaces, warning systems, or online driver monitoring systems \citep{manstetten2021}.

\subsection{Limitations and Future Research}
The present study suffers from several limitations. First, the study was performed on a closed test track. Even though test tracks provide a dynamic real-world environment which is controllable and safe, its complexity remains low. We suggest testing a similar paradigm in real traffic once the legislative and technological limitations are overcome. 

Moreover, our data seems to have relatively low sensitivity, likely originating from environmental noise. We recommend recruiting larger sample sizes, using high-density EEG systems, and keeping the impedance below $10 k\Omega$ for EEG data collection in the wild.

Additionally, this study was relatively short in duration (17 minutes per each condition). Some aspects of system use facilitated by automation, such as the need for the physical operation of controls, may only play a role during long-term or repeated usage. We recommend conducting longer drives and/or repeated measurement points to observe the dynamics in the drivers' experience. Moreover, longer periods of driving would provide more data and could therefore improve the signal-to-noise ratio.

The study was not designed to investigate the role of experience with automation on perceived cognitive load and attentional resource allocation. Nonetheless, some participants mentioned that the system's novelty could have had an impact, which could change with long-term exposure. Future studies should investigate this potential relationship and verify whether it exists. Moreover, it would be interesting to compare young drivers and older drivers. We also suggest addressing the spatial localization of the stimuli on the stimuli processing and addressing the crossmodal potential to improve the driver-vehicle interaction. 

Future investigations could also add to the interpretation of overt driver attention by measuring driver gaze behaviour \citep{bieg2020task} with an eye-tracking system in conjunction with attentional measures based on EEG. Additionally, it would be interesting to use eye tracking for detecting how relevant environmental cues (e.g., bicycle ring, ambulance siren) prime drivers to pay more attention to relevant visual scenes.

Finally, future studies should explore the effect of top-down attention allocation on the processing of environmental stimuli. In the present study, we could not reliably distinguish the difference between the effect of attention allocation based on cognitive goals and task demands that would imply competing for resources. Moreover, we recommend addressing the tonic changes in alpha and theta frequency bands to gain more insights into fatigue and underload processes in future studies.  

\section{Conclusion}
This study provides new empirical evidence about the cognitive mechanisms underlying driver-vehicle interactions in a realistic environment. We studied attentional resource allocation and subjective cognitive load during manual, SAE L2, and SAE L3 driving. Our results suggest that drivers perceive L3 driving as the least demanding, while no substantial difference in self-reported subjective cognitive load is observed between manual and L2 driving. Moreover, we studied the P3a ERP component elicited by the passive auditory oddball task. The amplitude of the P3a ERP component indicates the amount of cognitive resources invested into processing the auditory stimuli. It can serve as an objective indicator of cognitive load and attention allocation. We found diminished P3a amplitudes in both L2 and L3 driving compared to manual driving. This may suggest that during automated driving, drivers allocate fewer attentional resources to processing environmental information. However, further empirical evidence is needed to understand whether these differences can be attributed to top-down attention control leading to attention withdrawal or bottom-up competition for resources induced by cognitive load. The effect of other factors, such as experience and novelty, should also be addressed in future studies. To ensure the safety, efficiency, and comfort of future AVs, it is crucial to have a comprehensive understanding of how drivers allocate their attentional resources when operating an automated vehicle and the effects of automation on cognitive load. Despite its limitations, our study offers important insights into real-world driver-vehicle interactions. It highlights the importance of managing drivers’ attention and cognitive load with implications for enhancing automation safety and the design of user interfaces. 

\section*{Conflict of Interest Statement}
The authors declare that the research was conducted in the absence of any commercial or financial relationships that could be construed as a potential conflict of interest.

\section*{Author Contributions}
Conceptualization: NF, HJB, LC. Methodology: NF, HJB, LC. Software: HJB, NF. Formal analysis: NF, HJB, LC. Investigation: NF, HJB, YC. Resources: OP, HJB, MB. Data curation: NF, JER. Original draft: NF, JER. Review and editing: OP, HJB, JER, YC, LC, MB. Visualisation: NF. Supervision: OP, MB, LC. Project administration: HJB. Funding acquisition: OP, HJB, MB. 

\section*{Funding}
This project has received funding from the European Union's Horizon 2020 research and innovation programme under the Marie Sklodowska-Curie grant agreement 860410 and was supported by the German Federal Ministry for Economic Affairs and Climate Action (Bundesministerium für Wirtschaft und Klimaschutz, BMWK) through the project RUMBA (projekt-rumba.de).

\section*{Acknowledgments}
We want to thank Philipp Alt and Erdi Kenar from Robert Bosch GmbH and Jürgen Pichen from Ulm University for their involvement in the project.

\section*{Data Availability Statement}
The data can be found at https://doi.org/10.17632/yyndyf8wxb.1

\bibliographystyle{elsarticle-harv}
\bibliography{references}

\end{document}